\documentclass{JHEP}
\usepackage{epsfig}

\newcommand{\be}{\begin{equation}}
\newcommand{\ee}{\end{equation}}
\newcommand{\bea}{\begin{eqnarray}}
\newcommand{\eea}{\end{eqnarray}}
\newcommand{\bean}{\begin{eqnarray*}}
\newcommand{\eean}{\end{eqnarray*}}
\newcommand{\ba}{\begin{array}}
\newcommand{\ea}{\end{array}}

 \newcommand{\IR}{{\bf R}}

\def\beq{\begin{equation}}

\def\bra#1{\left\langle #1\right|}
\def\eeq{\end{equation}}

\def\ket#1{\left| #1\right\rangle}

\def\Arcsech{{\rm arcsech}}
\def\eref#1{(\ref{#1})}

\preprint{MIT-CTP-3254\\ {\tt hep-th/0203175}}

\title{Zeeman Spectroscopy of the Star Algebra}
\author{Bo Feng, Yang-Hui He and Nicolas Moeller
\\
Center for Theoretical Physics,
\\Massachusetts Institute of Technology,\\ Cambridge, MA 02139, USA\\
\email{fengb,yhe,moeller@ctp.mit.edu}
}

\abstract{We solve the problem of finding all eigenvalues and 
eigenvectors of the Neumann matrix of the matter sector of open
bosonic string  
field theory, including the zero modes, and switching on a background 
$B$-field. We give the discrete eigenvalues as roots of transcendental
equations,  
and we give analytical expressions for all the eigenvectors.}

\keywords{String Field Theory, Neumann coefficients, $B$-field}
\begin{document}

\section{Introduction}
After Rastelli, Sen and Zwiebach calculated, in \cite{Spec}, the spectrum and 
all the eigenvectors of the zero-momentum Neumann matrix in the 
matter sector $M$, it has become 
clear that the knowledge of the spectrum is very useful to perform exact 
calculations in string field theory \cite{Spec_Oku_1, Spec_Oku_2, Okuda_1}. 
Also, it has recently been shown by Douglas, Liu, Moore and Zwiebach in 
\cite{Moyal} that this allows us to write the 
star-product of two zero-momentum string fields as a continuous tensor product 
of Moyal products, each of which corresponding to one eigenvalue in
the spectrum  
of $M$. The noncommutativity parameter $\theta$ of each of these Moyal
products is  
given as a function of the eigenvalue $\lambda$. For $\lambda =
-{1\over3}$, it is  
zero, and we therefore have one commutative product corresponding to
the momentum  
carried by half of a string (which is conserved in this setup).

Because of this success, it has become a subject of interest to diagonalize 
Neumann matrices in diverse situations: In \cite{Super}, Mari\~no and 
Schiappa diagonalized the Neumann matrices of superstring field theory. In a 
recent paper \cite{mprime} we diagonalized $M'$, the bosonic Neumann matrix 
in the matter 
sector including zero modes; the same work was done independently by 
Belov \cite{Belov}. We found that the spectrum of $M'$ has the same
continuous part as that of $M$, that is the interval $[-{1\over3},
0)$, plus one doubly  
degenerate eigenvalue in the range $(0,1)$, whose value depends on the
parameter  $b$, defined by $a_0 = {1\over2} \sqrt{b} \hat{p} - {1
\over \sqrt{b}} i \hat{x}$. 

The goal of this paper is to diagonalize the Neumann matrix ${\cal M}^{11}$ 
with zero modes in a nontrivial $B$-field background. The form 
of the star-product in the presence of a $B$-field was already studied in 
\cite{Sugino,Kawano_B,Witten_B,Schnabl_B,Bonora}. 
We will use here the formalism of 
Bonora, Mamone and Salizzoni \cite{Bonora} for the Neumann matrix
${\cal M}^{11}$  with 
nonzero $B$-field, which we can easily cast into the form introduced
by Okuyama  \cite{Spec_Oku_2}. 
Therefore, it is merely an extension of our work \cite{mprime} 
to give complete expressions for the eigenvalues and eigenvectors of 
${\cal M}^{11}$.

In short, our results are as follows: We turn on a $B$-field 
$\left( \begin{array}{cc} 0 & B \\ -B & 0 \end{array} \right)$ in two
spatial  directions. First we find a scaling parameter $\xi = 1 + (2
\pi \alpha' B)^2$,  such that  
our continuous spectrum is $[-{1\over 3 \xi}, 0)$, the same as that of
$M$ or  $M'$ but shrunk by $\xi$. Each eigenvalue in this interval is
four times  
degenerate (they are twice more degenerate than the eigenvalues of
$M'$ simply because we are considering two spatial directions instead
of a single one), 
except the point $\lambda= -{1\over 3 \xi}$ which is only doubly
degenerate when $B \neq 0$. 
Then we find {\em two} doubly degenerate discrete eigenvalues in the
range  $(0, {1\over \xi})$ (again the same range as for $M'$ but
shrunk by $\xi$).  
We give these two eigenvalues in terms of roots of transcendental equations 
depending on $b$ and $B$. 
When $B=0$, these eigenvalues are the 
same, whereas the role of turning on a $B$-field is just to split them
and push one eigenvalue towards $0$ and the 
other one towards ${1\over \xi}$ (This is reminiscent of the 
Zeeman effect). Finally we give analytical expressions 
for all the eigenvectors.

This paper is organized as follows. We begin with nomenclature and
some review of the known results in diagonalising the Neumann matrix,
especially with the zero modes, in Section 2. Then in Section 3 
we proceed to the
basic setup of diagonalising the matrix in the presence of a $B$-field
background. Here we reduce the problem to the solution of a $4 \times
4$ linear system. We next address the case of the continuous spectrum
in Section 4. Thereafter we focus on the properties of the determinant
of the $4 \times 4$ system in Section 5 so as to obtain a discrete
spectrum in Section 6. To verify our analysis, we perform level
truncation analysis in Section 7 and come to satisfying agreement.
We end with conclusions and prospects in Section 8.

\section{Notations and Some Review of Known Results}

In a previous work \cite{mprime}, we generalised the results of
\cite{Spec,Spec_Oku_1,Spec_Oku_2} in studying the spectrum of the
Neumann Matrix by including the zero mode.
We recall that this is the matrix
\[
M':= \left( \begin{array}{cc} M'_{00} & M'_{0m} \\
  M'_{n0} & M'_{nm} \end{array} \right)=\left( \begin{array}{ll}
  1-\frac{2}{3} \frac{b}{ \beta} & 
  -\frac{2}{3} \frac{\sqrt{2b}}{ \beta} \bra{v_e}~~~~~~~ \\
  -\frac{2}{3} \frac{\sqrt{2b}}{ \beta} \ket{v_e} \qquad &
  M+\frac{4}{3} \frac{(-\ket{v_e}\bra{v_e}+\ket{v_o}\bra{v_0})}{ \beta} 
 \end{array} \right),
\]
with
\[
\beta := V_{00}^{rr}+\frac{b}{2} = \ln \frac{27}{16} + \frac{b}{2}, \qquad
\ba{l}
\ket{v_e} := E^{-1}\ket{A_e} \\
\ket{v_o} := E^{-1}\ket{A_o},
\ea
\quad E_{mn} :=\sqrt{n} \delta_{mn}
\]
and
\[
\ba{l}
(A_e)_n := \frac{1+(-)^n}{2} A_n \\
(A_o)_n := \frac{1-(-)^n}{2} A_n,
\ea
\quad \mbox{s.t.}~~(\frac{1+ix}{1-ix})^{1/3} := \sum_{n=even}
	A_n x^n+ i\sum_{n=odd}A_n x^n.
\]

Now we recall that
$M'$ has a continuous spectrum in $[-1/3,0)$ with eigenvalue $M(k) =
-\frac{1}{1+2\cosh \frac{\pi k}{2}}$, with a pair of degenerate
twist-even and twist-odd eigenvalues under the twist operator $C_{mn} :=
(-1)^m \delta_{mn}$. Moreover, $M'$ has an isolated eigenvalue inside
$(0,1)$ also with doubly degenerate eigenvectors.

We shall also adhere to the following definitions, as in
\cite{mprime}.
All the vectors will ultimately be written in the basis
\[
\ket{k}=(v_1^{k},v_2^{k},v_3^{k},...)^T
\]
with the generating function
\[
f_k(z)=\sum_{n=1}^{+\infty} \frac{v_n^{k}}{\sqrt{n}} z^n
=\frac{1}{k}(1-e^{-k\arctan z}).
\]

Defining $|z\rangle = (z,z^2,z^3,...)^T$, and the inner product
\[
\langle z\ket{k}\equiv \sum_{n=1}^{+\infty} z^n v_n^{k},
\]
we have orthonormality and closure conditions:
\[
\langle k | p \rangle= {\cal N}(k)\delta(k-p), \qquad 
{\cal N}(k) := \frac{2}{k} \sinh(\frac{\pi k}{2}); \quad
{\bf 1}=\int_{-\infty}^{+\infty} dk \frac{\ket{k}\bra{k}}{{\cal N}(k)}.
\]
Moreover under the twist action,
\[
C\ket{z}=\ket{-z}, \qquad C\ket{k}=-\ket{-k}.
\]

The $\ket{v_e}$ and $\ket{v_o}$ vectors above obey
\[
\bra{k}v_e\rangle=\frac{1}{k} \frac{\cosh(\frac{\pi k}{2})-1}
  {2\cosh(\frac{\pi k}{2})+1},~~~~\bra{k}v_o \rangle=\frac{\sqrt{3}}{k} 
\frac{\sinh(\frac{\pi k}{2})}  {2\cosh(\frac{\pi k}{2})+1}
\]
and
\[
C\ket{v_e}=\ket{v_e}, \qquad C\ket{v_o}=-\ket{v_o}.
\]

Finally, with the inner product the generating functions can be
written as
\[
f_k(z)=\bra{z}E^{-1}\ket{k}=\langle k|E^{-1}\ket{z},
\]
and we define also 
\[
G_e(z)\equiv\bra{z} E^{-1} \frac{1}{\lambda-M} \ket{v_e} =
\int_{-\infty}^{+\infty} 
  dk { f_k(z) \bra{k} v_e \rangle \over {\cal N}(k)(\lambda-M(k)) },
\label{gen_ve}
\]
\[
G_o(z)\equiv\bra{z} E^{-1} \frac{1}{\lambda-M} \ket{v_o} =
  \int_{-\infty}^{+\infty} 
  dk { f_k(z) \bra{k} v_o \rangle \over {\cal N}(k)(\lambda-M(k)) },
\]
whose explicit integrations were carried out in Equations (7.6) and
(7.10) of \cite{mprime}.
\subsection{The Presence of the Background $B$-Field}
Now let us consider the presence of a background field
$B^{\alpha \beta}$. 
We use the notation in \cite{Bonora}. The authors define the following
matrices 
\be
\label{bon_2.4}
 G^{\alpha \beta} =  ( {1 \over \eta+ 2\pi \alpha' B} \eta
 {1 \over \eta- 2\pi \alpha' B})^{\alpha \beta} ,~~~~\theta^{\alpha \beta}  
= - (2 \pi \alpha')^2( {1 \over \eta+ 2\pi \alpha' B} B {1 \over \eta- 
2\pi \alpha' B})^{\alpha \beta},
\ee
where they have chosen the simplest but easily generalizable case of
$B$ non-vanishing in two directions ${\alpha \beta}$ equaling to, say
24 and 25, and $\eta^{\alpha \beta}$ is the flat metric in
 $\IR^2$. 

Adhering to their convention, we explicitly chose
\be
\label{bon_2.15} B_{\alpha \beta}=\left( \begin{array}{cc} 0 & B \\ -B & 0 
  \end{array} \right),~~~~~~\eta_{\alpha \beta}=\left( \begin{array}{cc}
  1 & 0 \\ 0 & 1 \end{array} \right).
\ee
From \eref{bon_2.15} we can simplify \eref{bon_2.4} to
\be
\label{G_theta} G^{\alpha \beta}= {1 \over \xi}\left( \begin{array}{cc}
 1 & 0 \\ 0 & 1 \end{array} \right),~~~~~
 \theta^{\alpha \beta}\equiv \theta \epsilon^{\alpha \beta}= 
  { -(2\pi \alpha')^2 B \over \xi}\left( \begin{array}{cc}
  0 & 1 \\ -1 & 0 \end{array} \right),
\ee
where we have defined
\[
\xi= 1+ (2\pi \alpha' B)^2,~~~~\theta={-( 2\pi \alpha')^2 B \over \xi}.
\]
These immediately give us
\[
\det(G)=\xi^2, \qquad \theta \sqrt{\det(G)}=-( 2\pi \alpha')^2 B.
\]

For later usage and recalling $\beta := V_{00}^{rr}+{b\over 2}$, we
define two matrices
\be
\label{bon_2.9} \chi^{rs}=\left( \begin{array}{ccc} 0 & 1 & -1 \\ -1 & 0 
  & 1 \\ 1 & -1 & 0 \end{array} \right),~~~~~~
\phi^{rs}=\left( \begin{array}{ccc}  1 & -{1\over 2} & -{1\over 2} \\
-{1\over 2} & 1 & -{1\over 2}  \\ -{1\over 2} & -{1\over 2} & 1
\end{array} \right),
\ee
as well as two parametres
\be
\label{omegaxi}
\Omega={ 2\beta \over 4 \pi^4 {\alpha'}^2 B^2 + 3 \beta^2 },~~~~~
\Xi={ i (\pi \alpha')^2 B \over \beta \xi}.
\ee
We shall henceforth take $\alpha' = 1$.

Using these notations we can write the Neumann coefficients in the
presence of the
background $B$-field \cite{Bonora}:
First the vertex is
\beq
|V_3\rangle = \int d^{26}p_{(1)} d^{26}p_{(2)} d^{26}p_{(3)} 
\delta^{(26)}(p_{(1)} + p_{(2)} + p_{(3)}) \exp(-E) |0,p \rangle_{123} \,,
\eeq
where $E = E_{\|} + E_{\perp}$ is split into a parallel part (for the
directions in which $B=0$) which is
the same as the vertex without $B$-field, and a perpendicular part 
(for the two spatial directions in which we switched on the $B$-field)  
which depends on $B$; it can be written
\beq
E_{\perp} = \sum_{r,s=1}^3 \left( {1\over2} \sum_{m,n\geq 1} G_{\alpha
\beta} a_m^{(r)\alpha \dagger} V_{mn}^{rs} a_n^{(s) \beta \dagger} + 
\sum_{n\geq 1}G_{\alpha \beta} p_{(r)}^\alpha V_{0n}^{rs} a_n^{(s)
\beta \dagger} + {1\over2} G_{\alpha \beta} p_{(r)}^\alpha V_{00}^{rs} 
p_{(s)}^\beta + {i\over2} \sum_{r<s} p_\alpha^{(r)} \theta^{\alpha
\beta} p_\beta^{(s)} \right) \,.
\label{Eperp}
\eeq
Here we notice that if we don't include the zero-modes in our
analysis, \eref{Eperp} would simply reduce to 
$$
\sum_{r,s=1}^3 \left( {1\over2} \sum_{m,n\geq 1} G_{\alpha
\beta} a_m^{(r)\alpha \dagger} V_{mn}^{rs} a_n^{(s) \beta \dagger}
\right) \,.
$$
We thus see immediately, from the form of 
$G_{\alpha \beta}={1\over\xi}\eta_{\alpha \beta}$ in
\eref{G_theta}, that in this case, 
the effect of $B$ is just to shrink the spectrum by $\xi$.
We will thus include the zero-modes in our analysis. In this case, we
write the vertex $|V_3\rangle = |V_{3,\perp}\rangle \otimes |V_{3,
\|}\rangle$, where $|V_{3,\|}\rangle$ is the same as in the case
without $B$-field and $|V_{3,\perp}\rangle = K e^{-E} |\Omega\rangle$
where $|\Omega\rangle$ is annihilated by all $a_m$, $m \geq 0$, $K$ is
some constant which depends on $b$ and $B$ \cite{Bonora}, and $E$ is
$$
E = {1\over2} \sum_{r,s=1}^3 \sum_{m,n \geq0}
a_m^{(r)\alpha\dagger} {\cal V}_{mn}^{\alpha\beta,rs}
a_n^{(s)\beta\dagger} \,,
$$
where\footnote{In \ref{bon_2.23}, we have used
$-\Omega$ instead of $\Omega$ in \cite{Bonora} since when we set $B=0$
we should get back to the zero $B$-field case.} 
\bea
\label{bon_2.21} {\cal V}^{\alpha \beta,rs}_{00} & = & 
  G^{\alpha \beta} \delta^{rs} -\Omega b (G^{\alpha \beta} \phi^{rs}
  +\Xi\epsilon^{\alpha \beta} \chi^{rs}), \\
\label{bon_2.22} {\cal V}^{\alpha \beta,rs}_{0n} & = & \Omega \sqrt{b}
  \sum_{t=1}^3 (G^{\alpha \beta} \phi^{rt}+\Xi\epsilon^{\alpha \beta} 
\chi^{rt}) V_{0n}^{ts}, \\
\label{bon_2.23} {\cal V}^{\alpha \beta,rs}_{mn} & = & G^{\alpha \beta}
  V_{mn}^{rs} -\Omega \sum_{t,v=1}^3 V_{m0}^{rv}
 (G^{\alpha \beta} \phi^{vt}+\Xi\epsilon^{\alpha \beta} \chi^{vt})
 V_{0n}^{ts} \,.
\eea
These coefficients satisfy the properties
\[
{\cal V}^{\alpha \beta,rs}_{mn}={\cal V}^{\beta\alpha,sr}_{nm},~~~~~
{\cal V}^{\alpha \beta,rs}_{mn}={\cal V}^{\alpha \beta,(r+1)(s+1)}_{mn}
\,.
\]

To write these coefficients into a more
compact form, we need to invoke from Appendix B \cite{0102112} the
following forms for the Neumann Coefficients, rewritten in our basis: 
\[
\ba{cclccl}
V_{n0}^{rr} & = &  {-2 \sqrt{2} \over 3} \ket{v_e}, 
	& V_{0n}^{rr} & = & {-2 \sqrt{2} \over 3} \bra{v_e}, \\
V_{n0}^{21}& = &{\sqrt{2} \over 3}\ket{v_e} +{ \sqrt{6} \over 3} \ket{v_o}, &
V_{0n}^{12}& = &{\sqrt{2} \over 3} \bra{v_e} +{ \sqrt{6} \over 3}\bra{v_o},\\
V_{n0}^{12}& = &{\sqrt{2} \over 3} \ket{v_e}-{ \sqrt{6} \over 3}  \ket{v_o},&
V_{0n}^{21}& = & { \sqrt{2} \over 3} \bra{v_e} -{ \sqrt{6} \over 3}
\bra{v_o}.
\ea
\]

Substituting the above, \eref{bon_2.21}, \eref{bon_2.22} and \eref{bon_2.23}
simplify to
\bean
({\cal V}^{\alpha \beta}_{mn})^{11} & = & G^{\alpha \beta}[ V_{mn}^{11}
 - 2\Omega( \ket{v_e} \bra{v_e}+\ket{v_o} \bra{v_o})] \nonumber \\
 & - &
{ 4\over \sqrt{3}} \Omega \Xi \epsilon^{\alpha \beta}[\ket{v_e} \bra{v_o}-
\ket{v_o} \bra{v_e}], \\ 
({\cal V}^{\alpha \beta}_{0n})^{11} & = & -\sqrt{2b} \Omega
G^{\alpha \beta} \bra{v_e}-{2 \sqrt{6b} \over 3} 
 \Omega \Xi \epsilon^{\alpha \beta} \bra{v_o}, \\
({\cal V}^{\alpha \beta}_{n0})^{11} & = &({\cal V}^{\beta\alpha}_{0n})^{11}
  \nonumber \\
& = & -\sqrt{2b} \Omega G^{\alpha \beta} \ket{v_e}+{2 \sqrt{6b} \over 3} 
 \Omega \Xi \epsilon^{\alpha \beta} \ket{v_o}, \\
({\cal V}^{\alpha \beta}_{00})^{11} & = & G^{\alpha \beta} (1-\Omega b),
\eean

We can define two matrices
\bea
\Gamma&:=&\left( \begin{array}{ll} 1-\Omega b & -\sqrt{2b} \Omega\bra{v_e}\\
-\sqrt{2b} \Omega \ket{v_e}~~~& V^{11}- 
2\Omega( \ket{v_e} \bra{v_e}+\ket{v_o} \bra{v_o})
\end{array} \right)
\nonumber \\
\Sigma&:=&\left( \begin{array}{ll} 0 & -{2 \sqrt{6b} \over 3} \Omega \Xi \bra{v_o}
\\ {2 \sqrt{6b} \over 3}\Omega \Xi\ket{v_o}~~~ &
 -{ 4\over \sqrt{3}} \Omega \Xi[\ket{v_e} \bra{v_o}-
\ket{v_o} \bra{v_e}]  
\end{array} \right)
\label{gammasigma}
\eea
to simplify things further.

Using \eref{gammasigma} we can finally write down the Neumann coefficients
$({\cal V})^{11}$ in the presence of the $B$-Field into a
$4\times 4$ matrix form
\be
({\cal V})^{11}=\left( \begin{array}{ll} { 1\over \xi} \Gamma & \Sigma \\
-\Sigma & { 1\over \xi} \Gamma
\end{array} \right)  
\ee

\subsection{The Matrix of Our Concern: $({\cal M})^{11}$}

Combining all of the notation above, 
the matrix we wish to diagonalize is $({\cal V})^{11}$ multiplied by the
twist operator $C$, i.e.,
\be
\label{M_11}
({\cal M})^{11}\equiv C({\cal V})^{11}=\left( \begin{array}{ll} 
{ 1\over \xi} C\Gamma & C\Sigma \\
-C\Sigma & { 1\over \xi} C\Gamma
\end{array} \right)  
\ee
where
\[
C\Gamma := \left( \begin{array}{ll} 1-\Omega b & -\sqrt{2b} \Omega\bra{v_e}\\
-\sqrt{2b} \Omega \ket{v_e}~~~& M- 
2\Omega( \ket{v_e} \bra{v_e}-\ket{v_o} \bra{v_o})
\end{array} \right)
\]
and
\[
C\Sigma:=\left( \begin{array}{ll} 0 & -{2 \sqrt{6b} \over 3} \Omega \Xi \bra{v_o}
\\ -{2 \sqrt{6b} \over 3}\Omega \Xi\ket{v_o}~~~ &
 -{ 4\over \sqrt{3}} \Omega \Xi\left(\ket{v_e} \bra{v_o}+
\ket{v_o} \bra{v_e}\right)  
\end{array} \right).
\]
Note that since $\Xi$ from \eref{omegaxi} is a purely imaginary
number, $({\cal M})^{11}$ is Hermitian, so its eigenvalues are
real. It is the diagonalisation of this matrix  $({\cal M})^{11}$ with
which we shall concern ourselves in the remainder of the paper.

\section{Diagonalising  $({\cal M})^{11}$: The Setup}

Now we can solve the eigenvectors and eigenvalues, as what we did in
\cite{mprime}, with the ansatz
\[
v=\left( \begin{array}{c} g_1 \\ \int dk h_1(k) \ket{k}  \\  g_2  \\
  \int dk h_2(k) \ket{k} \end{array} \right)
\]
Acting on $v$ by \eref{M_11}, we can write the eigen-equation into four 
parts as
\bea
\lambda g_1 & = & {1-\Omega b \over \xi}g_1 - {\sqrt{2b} \Omega \over \xi}
  {\cal C}^{(1)}_e-{2 \sqrt{6b} \over 3} \Omega \Xi {\cal C}^{(2)}_o,
\nonumber \\
\int_{-\infty}^{+\infty} dk\lambda h_1(k) \ket{k} & = & 
 - {\sqrt{2b} \Omega \over \xi} \ket{v_e} g_1 +
  \int_{-\infty}^{+\infty} dk {M(k) \over \xi}h_1(k) \ket{k} 
   -  {2\Omega \over \xi} [ \ket{v_e}{\cal C}^{(1)}_e-\ket{v_o}  
  {\cal C}^{(1)}_o]
  \nonumber \\
 & - & {2 \sqrt{6b} \over 3} \Omega \Xi \ket{v_o} g_2 -
  { 4\over \sqrt{3}} \Omega \Xi[\ket{v_e} {\cal C}^{(2)}_o
+\ket{v_o} {\cal C}^{(2)}_e], \nonumber \\
\lambda g_2 & = &{2 \sqrt{6b} \over 3} \Omega \Xi {\cal C}^{(1)}_o
  +{1-\Omega b \over \xi}g_2- {\sqrt{2b} \Omega \over \xi} {\cal
C}^{(2)}_e, \nonumber \\
\int_{-\infty}^{+\infty} dk\lambda h_2(k) \ket{k} & = &
  {2 \sqrt{6b} \over 3} \Omega \Xi \ket{v_o} g_1+{ 4\over \sqrt{3}} 
  \Omega \Xi[\ket{v_e} {\cal C}^{(1)}_o+\ket{v_o} {\cal C}^{(1)}_e] 
  -  {\sqrt{2b} \Omega \over \xi} \ket{v_e} g_2 \nonumber \\
  & + & \int_{-\infty}^{+\infty} dk {M(k) \over \xi}h_2(k) \ket{k}- 
  {2\Omega \over \xi} [ \ket{v_e}{\cal C}^{(2)}_e-\ket{v_o}  
  {\cal C}^{(2)}_o]
\label{4eqs}
\eea
where we have defined
\bean
\label{C_eo}
{\cal C}^{(1)}_e & = &\int_{-\infty}^{+\infty} dk h_1(k)\bra{v_e}
k\rangle, \nonumber \\
{\cal C}^{(1)}_o & = & \int_{-\infty}^{+\infty} dk h_1(k)\langle v_o| k\rangle,
  \nonumber \\
{\cal C}^{(2)}_e & = &\int_{-\infty}^{+\infty} dk h_2(k)\bra{v_e} k\rangle, 
\nonumber  \\
{\cal C}^{(2)}_o & = & \int_{-\infty}^{+\infty} dk h_2(k)\langle v_o| k\rangle.
\eean
Now we solve $g_1,g_2$ from the first and third equations of
\eref{4eqs}:
\bean
g_1 & = & [- {\sqrt{2b} \Omega \over \xi}
  {\cal C}^{(1)}_e-{2 \sqrt{6b} \over 3} \Omega \Xi {\cal C}^{(2)}_o]/
  (\lambda-{1-\Omega b \over \xi}), \\
g_2 & = & [{2 \sqrt{6b} \over 3} \Omega \Xi {\cal C}^{(1)}_o- 
  {\sqrt{2b} \Omega \over \xi} {\cal C}^{(2)}_e]/
  (\lambda-{1-\Omega b \over \xi}).
\eean

Putting these back into the second and fourth equations of
\eref{4eqs} we obtain
\bea
\int_{-\infty}^{+\infty} dk h_1(k) \ket{k}[\lambda-{M(k) \over \xi}]
&=& \ket{v_e} [{\cal C}^{(1)}_e d_{ee}+{\cal C}^{(2)}_o d_{oe}]
  + \ket{v_o} [{\cal C}^{(1)}_o d_{oo}+ {\cal C}^{(2)}_e d_{oe}]  
\nonumber \\
\int_{-\infty}^{+\infty} dk h_2(k) \ket{k}[\lambda-{M(k) \over \xi}]
&=& \ket{v_o}[{\cal C}^{(2)}_o d_{oo}- {\cal C}^{(1)}_e d_{oe}]
+\ket{v_e}[ {\cal C}^{(2)}_e d_{ee} -{\cal C}^{(1)}_o d_{oe}], \qquad
\label{h1h2}
\eea
where to simplify notation, we have defined
\bean
d_{ee} & = & -{ 2\Omega(\xi \lambda-1 ) \over 
  \xi(\xi \lambda-1 +  \Omega b)}, \\
d_{oo} & = & { -2 \Omega \over 3 \xi}[-3+{ 4 b\Omega \Xi^2 \xi^2 \over
\xi \lambda-1 +  \Omega b }],\\
d_{oe} & = & -{ 4 \Omega \Xi 
 (\xi \lambda-1 ) \over \sqrt{3} (\xi \lambda-1 +  \Omega b)}.
\eean
We note that $d_{ee},d_{oo}$  are reals while $d_{oe}$ is purely
imaginary.

We can expand $\ket{v_e},\ket{v_o}$ as in (4.7) of \cite{mprime}
\[
\ket{v_e}=\int_{-\infty}^{+\infty} dk \ket{k}\frac{\bra{k}v_e
   \rangle}{{\cal N}(k)},\qquad 
\ket{v_o}=\int_{-\infty}^{+\infty} dk
   \ket{k}\frac{\bra{k}v_o\rangle}{{\cal N}(k)}.
\]
Subsequently, \eref{h1h2} can be re-written as\footnote{The term 
${ 1\over \lambda-{M(k) \over \xi}}$ is not very well defined when
we write it in this form. However, the only physically meaningful 
quantity is the expression $\int dk h(k) \ket{k}$. When we perform the
integration, as what we did in the generating function, we should
choose the principal-value integration. 
This fixes the definition. We want
to thank Dmitri Belov for discussing with us about this point.}
\bea
h_1(k) & = & { 1\over \lambda-{M(k) \over \xi}}[{\langle k|v_e\rangle
\over {\cal N}(k)}(d_{ee}  
  {\cal C}^{(1)}_e+d_{oe} {\cal C}^{(2)}_o)+{\langle k|v_o\rangle
\over {\cal N}(k)}( d_{oo}{\cal C}^{(1)}_o 
  + d_{oe} {\cal C}^{(2)}_e)+\delta(k-k_1) r_1(k)], \nonumber\\
h_2(k) & = & { 1\over \lambda-{M(k) \over \xi}}[{\langle k|v_e\rangle
\over {\cal N}(k)}(d_{ee}  
  {\cal C}^{(2)}_e-d_{oe} {\cal C}^{(1)}_o)+{\langle k|v_o\rangle
\over {\cal N}(k)}(d_{oo}{\cal C}^{(2)}_o 
  - d_{oe}{\cal C}^{(1)}_e)+\delta(k-k_2) r_2(k)]
\label{h1h2final}
\eea 
with yet undetermined parameters $k_1,k_2$ and functions
$r_1(k),r_2(k)$ with zeros respectively at $k_1,k_2$.
Again, as in \cite{mprime}, we define 
\bea
 A_{ee}(\lambda) &=&  \int_{-\infty}^{+\infty} dk
  \frac{\bra{k}v_e\rangle \langle v_e\ket{k}}{{\cal N}(k)
  (\lambda-{M(k) \over \xi})} = \bra{v_e} \frac{1}{\lambda - {M\over
\xi}} \ket{v_e}, \nonumber \\
 A_{eo}(\lambda) &=&  \int_{-\infty}^{+\infty} dk
  \frac{\bra{k}v_o\rangle \langle v_e\ket{k}}{{\cal N}(k)
  (\lambda-{M(k) \over \xi})}  = \bra{v_e} \frac{1}{\lambda - {M\over
\xi}} \ket{v_o}=0,  \nonumber \\
 A_{oo}(\lambda) &=&  \int_{-\infty}^{+\infty} dk
  \frac{\bra{k}v_o\rangle \langle v_o\ket{k}}{{\cal N}(k)
  (\lambda-{M(k) \over \xi})} = \bra{v_o} \frac{1}{\lambda - {M\over
\xi}} \ket{v_o} ,\nonumber \\
 B_{e}^{(1)}(\lambda) &=&  \int_{-\infty}^{+\infty} dk 
   \frac{\delta(k-k_1)  r_1(k)\langle v_e\ket{k}}{\lambda-{M(k) \over
\xi}}, \nonumber \\
 B_{o}^{(1)}(\lambda) &=&  \int_{-\infty}^{+\infty} dk 
   \frac{\delta(k-k_1)  r_1(k)\langle v_o\ket{k}}{\lambda-{M(k) \over
\xi}},\nonumber \\
 B_{e}^{(2)}(\lambda) &=&  \int_{-\infty}^{+\infty} dk 
   \frac{\delta(k-k_2)  r_2(k)\langle v_e\ket{k}}{\lambda-{M(k) \over
\xi}}, \nonumber \\
 B_{o}^{(2)}(\lambda) &=&  \int_{-\infty}^{+\infty} dk 
   \frac{\delta(k-k_2)  r_2(k)\langle v_o\ket{k}}{\lambda-{M(k) \over
\xi}}.
\label{ab}
\eea

Finally, we can write the eigen-equation \eref{4eqs} we wish to solve
into matrix form:
\be
\label{mateigen}
\left( \begin{array}{cccc}  1-d_{ee} A_{ee}~~ & 0 & 0 & -d_{oe} A_{ee} \\
  0 & 1-d_{oo} A_{oo}~~ & -d_{oe} A_{oo} & 0 \\
  0 &  d_{oe} A_{ee}  & 1- d_{ee} A_{ee}~~ & 0 \\
  d_{oe} A_{oo} & 0 & 0 & 1-d_{oo} A_{oo}
\end{array} \right)
\left( \begin{array}{l} {\cal C}^{(1)}_e \\ {\cal C}^{(1)}_o \\
  {\cal C}^{(2)}_e \\{\cal C}^{(2)}_o  \end{array} \right)
=\left( \begin{array}{l} B^{(1)}_e \\ B^{(1)}_o \\
  B^{(2)}_e \\  B^{(2)}_o  \end{array} \right).
\ee
To solve this equation we need to consider the determinant
$Det$ of the $4 \times 4$ matrix in \eref{mateigen}. We will leave this
discussion to section \ref{sec:det}. For now let us address the case
when $Det \ne 0$ so that we can invert \eref{mateigen}; this gives us
the continuous spectrum.
\section{The Continuous Spectrum}

As we will see from Section \ref{sec:det},
only a few $\lambda$'s make the determinant zero.
For other $\lambda$, the determinant is non-zero and we can invert to
solve 
${\cal C}$'s. Just as what we did in \cite{mprime}, to get the nonzero
solution, $\lambda$ must be in the region ${1 \over
\xi}[-1/3,0)$. This is what is going to give us
the {\bf continuous spectrum}. 
\subsection{The Continuous Eigenvalues}
To see the above discussion more clearly, let us write down 
the explicit form of $B_e$ from \eref{ab}:
\[
B_e^{(1)}  =   \int_{-\infty}^{+\infty} dk \frac{1}{k} 
\frac{\cosh(\frac{\pi k}{2})-1}{2\cosh(\frac{\pi k}{2})+1}
  \frac{\delta(k-k_1)  r_1(k)}{\lambda-{M(k) \over \xi}},
\]
Since $r_1(k_1)=0$, the above integration will be zero unless the
denominator $\lambda-{M(k) \over \xi}$ also has a zero at
$k=k_1$:
\[
\lambda-{M(k_1) \over \xi}=0 \,.
\] 
Because $M(k_1) \in [-1/3,0)$ for any $k_1$, 
we know immediately that we have a
continuous spectrum for any
\be
\label{lambdacont}
\lambda = {M(k) \over \xi} \in {1 \over \xi}[-1/3,0);
\ee
this is our continuous eigenvalue.
Comparing with the result in \cite{mprime} we see that in the
background of $B$, the continuous spectrum is simply scaled by a
factor of ${1 \over \xi}$. 

\subsection{The Continuous Eigenvectors}

Now let us construct the eigenvector for the given $\lambda$ from
\eref{lambdacont} but not equal to ${-1 \over 3\xi}$. 
We will consider this
special point in Subsection \ref{sec:special}.

Now let us set
\be
\label{lambdak0}
\lambda={M(k_0) \over \xi} = -\frac{1}{\xi}\frac{1}{1+2\cosh \frac{\pi
	k_0}{2}}
\ee
for our eigenvalue from \eref{lambdacont}.
we expand $\lambda-M(k)/\xi$ around $k_0$ as
\bean
\lambda-{M(k)\over \xi}& = & {1 \over \xi} [M(k_0)-M(k)]={1 \over
\xi}[-\frac{dM}{dk}|_{k_0} (k-k_0)-\frac{1}{2} 
  \frac{d^2M}{dk^2}|_{k_0}(k-k_0)^2+....]   \\
  & = &{1 \over \xi}[ -\frac{\pi\sinh\frac{\pi k_0}{2} }{(1+2\cosh \frac{\pi k_0}{2})^2}
  (k-k_0)-\frac{1}{2} \frac{\pi^2+\frac{\pi^2}{2}\cosh \frac{\pi k_0}{2}
  -\pi^2 \sinh^2\frac{\pi k_0}{2} }{(1+2\cosh \frac{\pi k_0}{2})^3}
  (k-k_0)^2+.... ] \nonumber
\eean
Recall that we have two independent pairs of parameters
$(k_1,r_1(k))$ and $(k_2,r_2(k))$ with $r_i(k_i)=0$. 
We freely choose\footnote{Here we choose $k_1=k_2=k_0$ for convenience.
	We can equally choose $k_1=k_0,k_2=-k_0$. However, it is easy
	to see that the final result is same.}
$k_1=k_2=k_0$, $r_1(k)=D_1 \cdot (k-k_0)$ and
$r_1(k)=D_2 \cdot (k-k_0)$, where $D_{i=1,2}$ are arbitrary constants.
Then we have
\bea
\label{B12oe} 
B_e^{(1)} & = & -\frac{\xi D_1 (\cosh(\frac{\pi k_0}{2})-1)
   (2\cosh(\frac{\pi k_0}{2})+1)}{\pi k_0 \sinh\frac{\pi k_0}{2}} :=
  B_{e,k_0} D_1, \nonumber \\
B_o^{(1)} & = & - \sqrt{3} \frac{\xi D_1 (2\cosh(\frac{\pi
k_0}{2})+1)}{\pi k_0} := B_{o,k_0} D_1, \nonumber \\
B_e^{(2)} & = & -\frac{\xi D_2 (\cosh(\frac{\pi k_0}{2})-1)
   (2\cosh(\frac{\pi k_0}{2})+1)}{\pi k_0 \sinh\frac{\pi k_0}{2}} :=
 B_{e,k_0} D_2, \nonumber \\
B_o^{(2)} & = & - \sqrt{3} \frac{\xi D_2 (2\cosh(\frac{\pi
k_0}{2})+1)}{\pi k_0} := B_{o,k_0} D_2. 
\eea
From this we can solve
\bean
\left( \begin{array}{l} {\cal C}^{(1)}_e \\ {\cal C}^{(1)}_o \\
  {\cal C}^{(2)}_e \\{\cal C}^{(2)}_o  \end{array} \right)
 & = & {1 \over \Delta} 
 \left( \begin{array}{l} (1-d_{oo} A_{oo})B_e^{(1)} + d_{oe}A_{ee}B_o^{(2)}
  \\ (1-d_{ee} A_{ee})B_o^{(1)}+d_{oe} A_{oo} B_e^{(2)} \\
  -d_{oe}A_{ee}B_o^{(1)}+(1-d_{oo} A_{oo}) B_e^{(2)} \\
  -d_{oe} A_{oo}B_e^{(1)} +  (1-d_{ee} A_{ee}) B_o^{(2)}
\end{array} \right) \nonumber \\\label{Solu}
& = & {D_1 \over \Delta} 
 \left( \begin{array}{l} (1-d_{oo} A_{oo})B_{e,k_0} 
  \\ (1-d_{ee} A_{ee})B_{o,k_0} \\ -d_{oe}A_{ee}B_{o,k_0}\\
 -d_{oe} A_{oo}B_{e,k_0} \end{array} \right)+
{D_2 \over \Delta} 
 \left( \begin{array}{l}d_{oe}A_{ee} B_{o,k_0} \\ d_{oe} A_{oo}B_{e,k_0} \\
 (1-d_{oo} A_{oo}) B_{e,k_0} \\ (1-d_{ee} A_{ee}) B_{o,k_0}
 \end{array} \right)
\eean
where we have defined 
$$
\Delta := (1-d_{ee} A_{ee})(1-d_{oo} A_{oo})+d_{oe}^2 A_{ee} A_{oo}
$$
so that $\Delta^2 = Det$.

Now recalling that we can solve the $h_i$ from \eref{h1h2final} from
the ${\cal C}$'s and the $g_i$ as well; whence substituting back into
\eref{4eqs}, we obtain the eigenvector.
We see that 
the general solutions are just the linear combinations
of the $D_1$ and $D_2$ terms:
\bea
v(k_0)& = & { D_1 \over \Delta}\left( \begin{array}{l} 
- {\sqrt{2b} \Omega  \over (\lambda \xi -1 +\Omega b)}
 (1-d_{oo} A_{oo})B_{e,k_0}+ 
  {2 \sqrt{6b}  \Omega \Xi  \xi \over 3(\lambda\xi -1 +\Omega b)} 
d_{oe} A_{oo}B_{e,k_0} \\  \\
 {B_{e,k_0} (d_{ee}-(d_{ee}d_{oo}+d_{oe}^2)A_{oo}) 
\over \lambda-{M \over \xi}} \ket{v_e} 
+{B_{o,k_0} (d_{oo}-(d_{ee}d_{oo}+d_{oe}^2)A_{ee})
\over \lambda-{M \over \xi}} \ket{v_o} -{\xi\Delta \over\frac{dM}{dk}|_{k_0}} 
\ket{k_0} \\ \\
{2 \sqrt{6b}  \Omega \Xi  \xi \over 3(\lambda \xi-1 +\Omega b)}
(1-d_{ee} A_{ee})B_{o,k_0}+ {\sqrt{2b} \Omega  \over
(\lambda\xi -1 +\Omega b)} d_{oe}A_{ee}B_{o,k_0}\\  \\
{- d_{oe} B_{o,k_0} \over \lambda-{M \over \xi}} \ket{v_e}
+{ -d_{oe} B_{e,k_0} \over \lambda-{M \over \xi}} \ket{v_o}
\end{array} \right)  + \nonumber \\ &&  \nonumber\\
& + & 
{ D_2 \over \Delta}\left( \begin{array}{l} - {\sqrt{2b} \Omega  \over
(\lambda\xi -1 +\Omega b)}d_{oe}A_{ee} B_{o,k_0}-{2 \sqrt{6b}  \Omega \Xi  \xi \over 3(\lambda\xi -1 +\Omega b)} (1-d_{ee} A_{ee}) B_{o,k_0}\\  \\
{ d_{oe} B_{o,k_0} \over \lambda-{M \over \xi}} \ket{v_e}
+{ d_{oe} B_{e,k_0} \over \lambda-{M \over \xi}} \ket{v_o}
\\  \\
{2 \sqrt{6b}  \Omega \Xi  \xi \over 3(\lambda\xi -1 +\Omega b)}d_{oe} A_{oo}B_{e,k_0}- {\sqrt{2b} \Omega  \over
(\lambda \xi-1 +\Omega b)}(1-d_{oo} A_{oo}) B_{e,k_0} \\ \\
{B_{e,k_0} (d_{ee}-(d_{ee}d_{oo}+d_{oe}^2)A_{oo}) 
\over \lambda-{M \over \xi}} \ket{v_e} 
+{B_{o,k_0} (d_{oo}-(d_{ee}d_{oo}+d_{oe}^2)A_{ee})
\over \lambda-{M \over \xi}} \ket{v_o} -{\xi \Delta \over\frac{dM}{dk}|_{k_0}} 
\ket{k_0}
\end{array} \right)\nonumber
\eea

Since for a given $\lambda$ there are two corresponding $k$-values in
solving \eref{lambdak0}, viz.,
$k_0$ and $-k_0$, we get two degenerate states $v(k_0),v(-k_0)$ (in fact, the
degeneracy is four because for each we have one corresponding to $D_1$
and another to $D_2$). Using
\[
B_{e,k_0}=B_{e,-k_0},~~~~~B_{o,k_0}=-B_{o,-k_0},
~~~~~~\frac{dM}{dk}|_{k_0}=-\frac{dM}{dk}|_{-k_0}
\]
we can write the eigenstates which are even and odd with respect to $k_0$:
\bea
v_{1,\lambda}={ v(k_0)+ v(-k_0) \over 2} & = & 
{ D_1 \over \Delta}\left( \begin{array}{l} 
- {\sqrt{2b} \Omega  \over (\lambda\xi -1 +\Omega b)}
 (1-d_{oo} A_{oo})B_{e,k_0}+ 
  {2 \sqrt{6b}  \Omega \Xi  \xi \over 3(\lambda \xi-1 +\Omega b)} 
d_{oe} A_{oo}B_{e,k_0} \\  \\
 {B_{e,k_0} (d_{ee}-(d_{ee}d_{oo}+d_{oe}^2)A_{oo}) 
\over \lambda-{M \over \xi}} \ket{v_e} 
 -{\xi \Delta \over 2 \frac{dM}{dk}|_{k_0}} 
(\ket{k_0}-\ket{-k_0}) \\ \\
0\\  \\
{ -d_{oe} B_{e,k_0} \over \lambda-{M \over \xi}} \ket{v_o}
\end{array} \right)  \nonumber \\ & &  \nonumber \\ 
&+&
{ D_2 \over \Delta}\left( \begin{array}{l} 0\\  \\
{ d_{oe} B_{e,k_0} \over \lambda-{M \over \xi}} \ket{v_o}
\\  \\
{2 \sqrt{6b}  \Omega \Xi  \xi \over 3(\lambda \xi-1 +\Omega b)}d_{oe}
A_{oo}B_{e,k_0}- {\sqrt{2b} \Omega  \over 
(\lambda \xi -1 +\Omega b)}(1-d_{oo} A_{oo}) B_{e,k_0} \\ \\
{B_{e,k_0} (d_{ee}-(d_{ee}d_{oo}+d_{oe}^2)A_{oo}) 
\over \lambda-{M \over \xi}} \ket{v_e} 
 -{\xi \Delta \over 2 \frac{dM}{dk}|_{k_0}} 
(\ket{k_0}-\ket{-k_0})
\end{array} \right)\nonumber \\
&\equiv& {D_1 \over \Delta} \, v^e_{1, \lambda} + 
{D_2 \over \Delta} \, v^o_{1, \lambda} \label{eigenc1}
\eea
and
\bea
v_{2,\lambda}=
{ v(k_0)- v(-k_0) \over 2} & = & 
{ D_1 \over \Delta}\left( \begin{array}{l} 
0 \\  \\
{B_{o,k_0} (d_{oo}-(d_{ee}d_{oo}+d_{oe}^2)A_{ee})
\over \lambda-{M \over \xi}} \ket{v_o} -{\xi \Delta \over 2\frac{dM}{dk}|_{k_0}} 
(\ket{k_0}+\ket{-k_0}) \\ \\
{2 \sqrt{6b}  \Omega \Xi  \xi \over 3(\lambda \xi-1 +\Omega b)}
(1-d_{ee} A_{ee})B_{o,k_0}+ {\sqrt{2b} \Omega  \over
(\lambda\xi -1 +\Omega b)} d_{oe}A_{ee}B_{o,k_0}\\  \\
{- d_{oe} B_{o,k_0} \over \lambda-{M \over \xi}} \ket{v_e}
\end{array} \right)  \nonumber \\ & &  \nonumber\\
& + & 
{ D_2 \over \Delta}\left( \begin{array}{l} - {\sqrt{2b} \Omega  \over
(\lambda\xi -1 +\Omega b)}d_{oe}A_{ee} B_{o,k_0}-{2 \sqrt{6b}  \Omega
\Xi  \xi \over 3(\lambda \xi-1 +\Omega b)} (1-d_{ee} A_{ee})
B_{o,k_0}\\  \\ 
{ d_{oe} B_{o,k_0} \over \lambda-{M \over \xi}} \ket{v_e}
\\  \\
0\\ \\
{B_{o,k_0} (d_{oo}-(d_{ee}d_{oo}+d_{oe}^2)A_{ee})
\over \lambda-{M \over \xi}} \ket{v_o} -{\xi \Delta \over 2\frac{dM}{dk}|_{k_0}} 
(\ket{k_0}+\ket{-k_0})
\end{array} \right) \nonumber \\
&\equiv& {D_1 \over \Delta} \, v^o_{2, \lambda} + 
{D_2 \over \Delta} \, v^e_{2, \lambda} \,. \label{eigenc2}
\eea
However, we see that $v_{i,\lambda}$ are neither twist-even nor
twist-odd under the $C$-operator. 
The twist-even (odd) states in the first direction of our 2-dimensions
chosen for the $B$-field
are mixed with the twist-odd (even) states in the second direction.
This behaviour is of course
a result of turning on the background $B$-field.

\subsection{The Special Case of $\lambda=-1/(3\xi)$}
\label{sec:special}

Now, as promised, we discuss the special case of $\lambda=-1/(3\xi)$
which was excluded from the above. For general 
$B\neq 0$, $\lambda=-1/(3\xi)$ does not make the determinant zero. Then we 
can solve the equation as before and find the coefficients ${\cal C}$ and
subsequently the eigenvector.

However, one thing is special for this point: this is that 
$\frac{dM}{dk}|_{k_0=0}$ when $k_0=0$. In another
words, we should choose $r(k)=D \cdot k^2$ instead of the linear
$r(k)=D \cdot (k-k_0)$.
Under such a choice, we have $B_{e,0}=0$ and $B_{o,0}=-6\sqrt{3}
\xi/\pi$. So the solution is
\bea
v(k_0)& = & {\xi D_1 \over \Delta}\left( \begin{array}{l} 
 0 \\  
 {B_{o,0} (d_{oo}-(d_{ee}d_{oo}+d_{oe}^2)A_{ee})
\over -1/3-M} \ket{v_o} -{36 \Delta \over \pi^2}
\ket{0} \\ 
{2 \sqrt{6b}  \Omega \Xi  \xi \over 3\xi(\lambda \xi-1 +\Omega b)}
(1-d_{ee} A_{ee})B_{o,0}+ {\sqrt{2b} \Omega  \over \xi
(\lambda\xi -1 +\Omega b)} d_{oe}A_{ee}B_{o,0}\\  
{- d_{oe} B_{o,0} \over -1/3-M} \ket{v_e}
\end{array} \right)  + \nonumber \\ & &  \nonumber\\
& + & 
{\xi D_2 \over \Delta}\left( \begin{array}{l} - {\sqrt{2b} \Omega
\over \xi
(\lambda\xi -1 +\Omega b)}d_{oe}A_{ee} B_{o,0}-{2 \sqrt{6b}  \Omega
\Xi  \xi \over 3\xi(\lambda\xi -1 +\Omega b)} (1-d_{ee} A_{ee})
B_{o,0}\\  \\ 
{ d_{oe} B_{o,0} \over -1/3-M} \ket{v_e}
  \\
0\\
{B_{o,0} (d_{oo}-(d_{ee}d_{oo}+d_{oe}^2)A_{ee})
\over -1/3-M} \ket{v_o} -{36 \Delta \over \pi^2}
\ket{0}
\end{array} \right). \nonumber
\eea

Now at $k_0=0$ we get only two independent eigenvectors
instead of the four as discussed at the end of the last subsection.
This may seem a little surprising because when $B=0$, we do have four
eigenvectors (since we have two spatial directions instead of one in
\cite{mprime}: two of $v_{+,-{1\over 3}}$ and two of $v_{-,-{1\over
3}}$). However, after a careful analysis we see that the two
eigenvectors $v_{+,-{1\over 3}}$ arise because when $B=0$, 
$\lambda=-1/3$ gives zero determinant.
But this is not the case here and we thus lose these two vectors.
More concretely, every point $\lambda$ in the continuous region
which makes the determinant zero will kill two eigenvectors while
adding another two eigenvectors. At $B=0$ case, the eigenvectors which
should be killed are not there, so the net effect is to add two
more eigenvectors.

\section{The Determinant}
\label{sec:det}
Having addressed the continuous, let us proceed with the discrete
spectrum. This corresponds to the few cases when $Det$ becomes
zero. Therefore in this section we shall determine the roots of $Det$.
We see that the determinant $Det$ can be written as
\be
Det(\lambda)=[(1-d_{ee} A_{ee})(1-d_{oo} A_{oo})+d_{oe}^2 A_{ee}
A_{oo}]^2, 
\ee
so that any root $\lambda$ of $Det$ will always be at least a double-root.

Now we have the evaluations of all the pieces of $Det$ from
\cite{mprime}; the reader is referred to the sections on $A_{ee}$ and
$A_{oo}$ therein. Using these results from equations (5.1) and the one
at the beginning of subsection 5.2 in {\em
cit. ibid.}, we have the following analytic expressions for $Det$. We
have set $x := \xi \lambda$. Then for $-\frac13 \le x < 0$,
\begin{eqnarray}
\label{det-in}
Det&=&
\left(b^2\,\left( 1 + 3\,x \right)  + 
  4\,b\,\left( -1 + 3\,x \right) \,\log (\frac{27}{16}) + 
  4\,\left( -1 + x \right) \,
   \left( 4\,B^2\,{\pi }^4 + 
     3\,{\log (\frac{27}{16})}^2 \right)\right)^{-2} \times  
\nonumber \\
&&\left\{
b^2\,\left( 1 + 3\,x \right)  - 
  4\,b\,\left( 1 + 3\,x \right) \,
   \left( 2\,\gamma + f(x) + 
     \log (16) \right)  + 
\nonumber \right. \\
&&
  4\,\left( 4\,B^2\,{\pi }^4\,\left( -1 + x \right)  + 
     4\,{\gamma}^2\,
      \left( 1 + 3\,x \right)  + 
     f(x)^2\,\left( 1 + 3\,x \right)  + 
     16\,\left( 1 + 3\,x \right) \,{\log (2)}^2 + 
\right.
\nonumber \\
&& 
\left. \left.
8\,f(x)\,\left( \log (2) + x\,\log (8) \right)
         + 4\,\gamma\,
      \left( 1 + 3\,x \right) \,
      \left( f(x) + \log (16) \right)  
\right)
\right\}^2
\nonumber\\
\end{eqnarray}
where $\gamma$ is the Euler constant,
\[
f(x) := \psi(g(x)) + \psi(-g(x)), \qquad 
g(x) := \frac{i}{2\pi}\Arcsech\left(-\frac{2x}{1+x}\right),
\]
and $\psi(x)$ is the {\em digamma function}. From this expression, we
see immediately two results: (1) When $B=0$, $x=-1/3$ makes the determinant
zero; (2) When $B\neq 0$, $x=-1/3$ is not a root of the determinant. 
This shows that when $B\neq 0$, the point at $x=-1/3$ has a 
different behavior.

On the other hand for $x$ not in the region, viz., $x <-\frac13$ or $x
\ge 0$, we have the form of $Det$ as in \eref{det-in}, but with $f(x)$
replaced by
\[
h(x) := \psi(-g(x)) + \psi(1+g(x)).
\]

\subsection{Zeros of $Det$ Between $-1/3$ and $0$}

We observe from \eref{det-in}, that $Det$ inside $[-\frac13,0)$ is
actually an algebraic equation in the transcendental function $f(x)$ and
$x$. Therefore we can solve $Det=0$ by bringing it into the form where
one side is $f(x)$ and the other, an algebraic function of $x$ (i.e., by
solving for $f(x)$); this gives us
\begin{equation}
\label{SolveIn}
2 f(x) = \frac{\mp 4\,B\,{\pi }^2\,{\sqrt{1 - x}}\,{\sqrt{1 + 3\,x}} - 
  \left( 1 + 3\,x \right) \,
   \left( -b + 4\,\gamma + \log (256)
     \right)}{1+3x}. 
\end{equation}
As an immediate check we can see that when $B \rightarrow 0$, $x =
-1/3$ is a solution; this is of course in perfect agreement with the
case seen in \cite{mprime}. Now for nonzero values of $B$, $x=-1/3$ is
no longer a solution.
In the interior region when $x \ne -1/3$, we can safely cancel the
$(1+3x)$ factor and the right-hand side simply becomes a (square-root)
hyperbola whose size is governed by $B$ and whose position is shifted
by $b$, i.e.,
\begin{equation}
\label{SolveIn_1} 2f(x)=\mp 4B \pi^2\sqrt{{1-x \over
1+3x}}-(-b+4\gamma
 +8\log 2)
\end{equation}
Now we have 2 cases, the one with the minus sign in front of $B$ in
\eref{SolveIn} and the other, with a plus sign. 

\subsection*{Case 1: $2f(x) = -4(\gamma+\log 4) + 
b - \frac{4\,B\,{\pi }^2\,{\sqrt{1 - x}}}{{\sqrt{1 + 3\,x}}}$}

Let us examine the left graph in Figure \ref{f:intersectIn}.
Indeed for sufficiently small $b$ the RHS
drops below the left and there are no solution. At a critical value
$b_0(B)$, the two curves touch at a single point (and no more because
the LHS is an increasing function whose derivative is also
increasing while the RHS is an increasing function whose
derivative is decreasing.) Above this critical point, there will
always be two solutions, which is the case we have drawn in the plot.
To find the critical point we equate the derivatives of both sides
which gives
\[
- \frac{i(1+3x)}{8 \pi^3 x}\left( \psi'(g(x)) - \psi'(-g(x))\right) = B
\]
From this equation we can solve $x_0(B)$ as a function of $B$.
Then we can find the critical value $b_0(B)$ by  substituting back into
the original intersection problem to yield
\[
2f(x_0(B)) = -4(\gamma+\log 4) + 
b - \frac{4\,B\,{\pi }^2\,{\sqrt{1 - x_0(B)}}}{{\sqrt{1 +
3\,x_0(B)}}}.
\]
The solution of this transcendental equation is then the critical
value $b_0(B)$ which we seek.

\subsection*{Case 2: $2f(x) = -4(\gamma+\log 4) + 
b + \frac{4\,B\,{\pi }^2\,{\sqrt{1 - x}}}{{\sqrt{1 + 3\,x}}}$}

For the other solution with $+$ in front of $B$ we refer to the right
plot of Figure \ref{f:intersectIn}. Here the situation is easier. The
RHS is always a decreasing function so there is always a solution
between $-1/3$ and 0.
\EPSFIGURE[h]{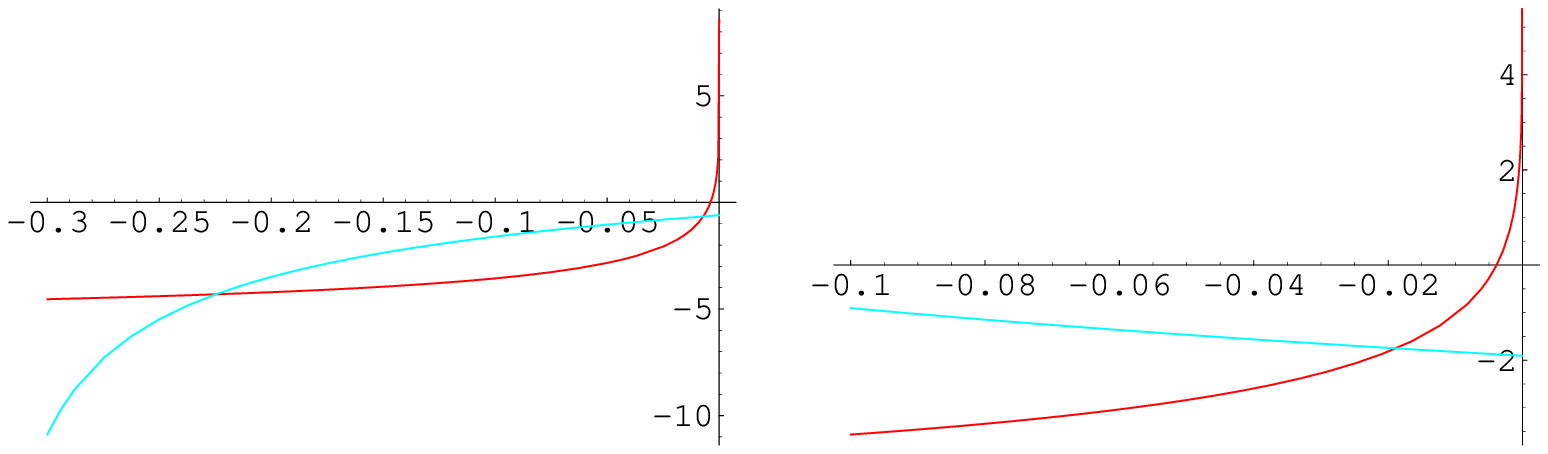,width=6in}
{The intersection showing the roots of $Det$ inside the region $x \in
(-1/3,0)$. The (red) convex
ascending curve in both cases is that of $2f(x)$.
In the graph to the left, we have chosen $b=11.2, B=0.1$ and have shown a
typical intersection of $2f(x)$ with the (blue) $-4(\gamma+\log 4) + 
b - \frac{4\,B\,{\pi }^2\,{\sqrt{1 - x}}}{{\sqrt{1 + 3\,x}}}$. There
could be $0,1$ or $2$ points of intersection depending on the shift
parametre $b$; the case shown is with 2 points of intersection.
On the graph to the right, we have chosen $b=2, B=0.1$ and the (blue)
curve is now $-4(\gamma+\log 4) + 
b + \frac{4\,B\,{\pi }^2\,{\sqrt{1 - x}}}{{\sqrt{1 + 3\,x}}}$. There
is always 1 point of intersection.
\label{f:intersectIn}
}

\subsubsection{Comparison Between the $B=0$ and $B\neq 0$ Cases}

Now let us compare the roots found here with the ones found in 
\cite{mprime} for the case $B=0$. There the determinant is reduced to
the diagonal form $\left| \begin{array}{ll} 1-d_{ee} A_{ee} & 0 \\ 0 & 
1-d_{oo} A_{oo} \end{array} \right|$. It has a root at $\lambda=-1/3$ 
where $1-d_{ee} A_{ee}=0$ but $1-d_{oo} A_{oo}\neq 0$. Furthermore,
there is a critical value $b_0=8\log 2$, above which we have doubly
degenerate roots where both $1-d_{ee} A_{ee}=0$ and $1-d_{oo} A_{oo}=
0$. 

In our general $B\neq 0$ case here. Firstly we have a root in the case 2
above no matter
what $b$ is. This root obviously corresponds to  $\lambda=-1/3$ 
at $B=0$. Secondly, there is a critical value $b_0(B)$, above which we
have two roots. These two roots obviously correspond to the
doubly degenerate roots when $B=0$. Therefore, we observe that when $B\neq 0$ the
doubly degenerate roots are split (Zeeman effect). This will be a general 
picture which we will meet again when we discuss the root in the
region $(0,1)$.

\subsection{Zeros of $Det$ between 0 and 1}
\label{zeros01}

We recall the form of $Det$ from \eref{det-in}. Here as in the
previous subsection we reduce the problem of finding its zeros to the
intersection of a transcendental function with an algebraic one:
\beq
2h(x) = b \mp \frac{4\,B\,{\pi }^2\,{\sqrt{1 - x}}}
   {{\sqrt{1 + 3\,x}}} - 
  4\,\left( \gamma + \log (4) \right)
\label{transc}
\eeq
We see immediately that for $x \notin [0,1]$, the function $h(x)$
is real while the RHS is complex (with non-vanishing imaginary part) whence
{\sl there are no solutions there}. We therefore focus on the solutions
between 0 and 1 where both sides become real. Again we must analyse
the minus and the plus cases.

\subsection*{Case 1: $2h(x) = b - \frac{4\,B\,{\pi }^2\,{\sqrt{1 - x}}}
{{\sqrt{1 + 3\,x}}} - 4\,\left( \gamma + \log (4) \right)$}

In the region $(0,1]$, the LHS is a monotonically decreasing function
while the RHS, a monotonically increasing one. At the endpoint
$x=1$, the RHS equals $b -4 (\gamma +\log 4)$
and the LHS is $-4 (\gamma +\log 4)$.
Therefore at the lower limit of $b=0$ there is a solution at
$x=1$. For any other value of positive $b$, the RHS gets shifted
upwards and there will always be a solution.
The situation is illustrated in the left part of Figure
\ref{f:intersectOut}.

\subsection*{Case 2: $2h(x) = b + \frac{4\,B\,{\pi }^2\,{\sqrt{1 - x}}}
{{\sqrt{1 + 3\,x}}} - 4\,\left( \gamma + \log (4) \right)$}

Now in the interesting region both functions are monotonically
decreasing, but the LHS, from $\infty$ and the RHS, from a
finite value.
Once again if $b=0$ there is an intersection at $x=1$, but now there
is another point of intersection in $(0,1)$. For positive $b$ however,
the $x=1$ endpoints no longer meet but the intersection at another
point in
between still exists.
\EPSFIGURE[h]{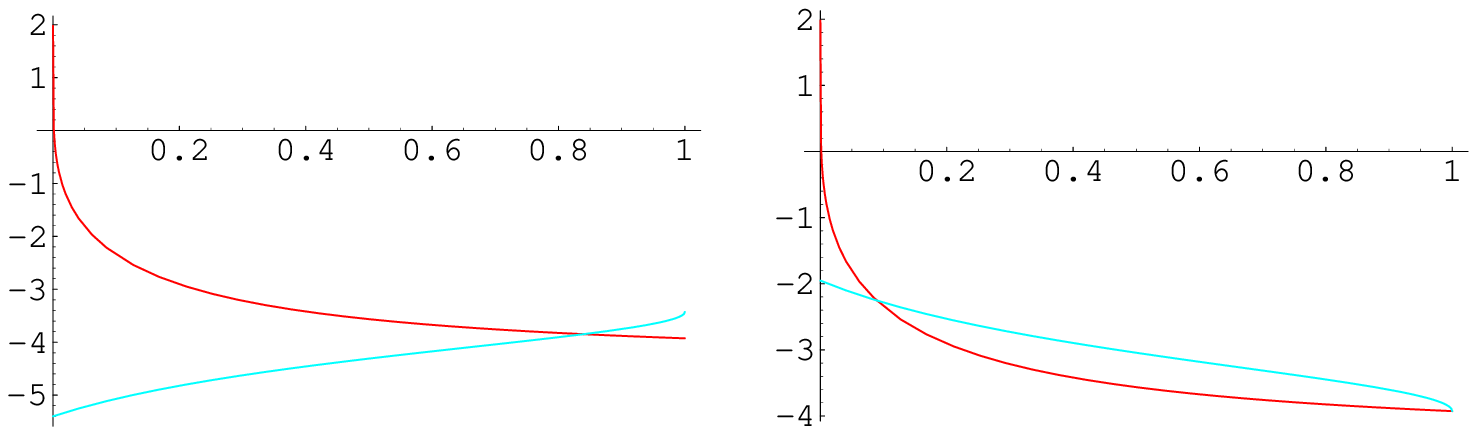,width=6in}
{The intersection showing the roots of $Det$ inside the region $x \in
(0,1]$. The (red) convex
descending curve in both cases is that of $2h(x)$.
In the graph to the left, we have chosen $b=1, B=0.1$ and have shown a
typical intersection of $2f(x)$ with the (blue) 
$b - \frac{4\,B\,{\pi }^2\,{\sqrt{1 - x}}}
{{\sqrt{1 + 3\,x}}} - 4\,\left( \gamma + \log (4) \right)$.
There is always 1 point of intersection.
On the graph to the right, we have chosen the limiting case $b=0,
B=0.1$ and the (red)
curve is now 
$b + \frac{4\,B\,{\pi }^2\,{\sqrt{1 - x}}}
{{\sqrt{1 + 3\,x}}} - 4\,\left( \gamma + \log (4) \right)$.
There is always 1 point of intersection in between and here there is
another at the endpoint $x=1$. For any other value of $b$, $x=1$ would
no longer be an intersection point.
\label{f:intersectOut}
}

Let us compare to the case of $B=0$ in \cite{mprime} again. 
There, we had
doubly degenerate roots in $(0,1)$. Here for general $B\neq 0$, we
have two roots which are not degenerate. It is obvious that these two
roots here correspond to the two doubly degenerate roots at 
$B=0$ and $B\neq 0$ lifts the degeneracy and splits the roots (Zeeman effect). 
The bigger the value of $B$ the further they split.
\section{The Discrete Spectrum}

Having addressed the case of the continuous spectrum corresponding to
$Det \ne 0$ in the preceding section, here we focus on the case
when $\lambda$ makes $Det$ zero. This will give us a {\bf Discrete
Spectrum}. 

\subsection{The Region ${ 1\over \xi}(0,1)$}

As we have seen before in this region, there are two values
$\lambda=\lambda_{1,2}$ making $Det=0$.
For both these values, $d_{oe} A_{ee}$ and $d_{oe} A_{oo}$ are nonzero
because they are zero only when $B=0$ or $\lambda=  {1\over \xi}$. 
Furthermore,
since $\lambda\in { 1\over \xi} (0,1)$, $\lambda-{M(k) \over \xi}$ can
not have a zero 
to cancel the ones from $r_1(k)$ and $r_2(k)$ in \eref{ab} and we get 
\[
B_{e}^{(1)}=B_{o}^{(1)}=B_{e}^{(2)}=B_{o}^{(2)}=0.
\]
This must indeed be so for \eref{mateigen} to have solutions for
$Det=0$. 

Now we can solve the ${\cal C}$'s, which must be in the nullspace of
the $4 \times 4$ matrix in \eref{mateigen}; we find two independent
vectors: 
\[
\left( \begin{array}{l} {\cal C}^{(1)}_e \\ {\cal C}^{(1)}_o \\
  {\cal C}^{(2)}_e \\{\cal C}^{(2)}_o  \end{array} \right)
=D_1\left( \begin{array}{l} d_{oe} A_{ee} \\ 0 \\ 0 \\ 1-d_{ee} A_{ee}
\end{array} \right)
+D_2\left( \begin{array}{l} 0 \\ 1-d_{ee} A_{ee} \\ -d_{oe} A_{ee}  \\0 
\end{array} \right)
\]
Substituting this back into \eref{h1h2final} and \eref{4eqs} as was
done before, we have:
\be
\label{discrete_1}
v  =  D_1 \left( \begin{array}{l} -{2 \sqrt{6b}  \Omega \Xi  \xi \over 
3(\lambda\xi -1 +\Omega b)}   \\
{ d_{oe} \over \lambda-{M \over \xi}} \ket{v_e} \\ 0 \\
{ d_{oo} - A_{ee} (d_{oo} d_{ee}+ d_{oe}^2) 
\over \lambda-{M \over \xi}} \ket{v_o} \end{array} \right)+
D_2 \left( \begin{array}{l} 0 \\ { d_{oo} - A_{ee} (d_{oo} d_{ee}+ d_{oe}^2) 
\over \lambda-{M \over \xi}} \ket{v_o} \\
{2 \sqrt{6b}  \Omega \Xi  \xi \over 
3(\lambda\xi -1 +\Omega b)}
\\ -{ d_{oe} \over \lambda-{M \over \xi}} \ket{v_e}
\end{array} \right)
\ee
where $\lambda=\lambda_{1,2}$, the two solutions of $Det$ in ${ 1\over
\xi}(0,1)$ given in the previous section.
\subsection{The Region $ { 1\over \xi} [-1/3,0)$}
When $\lambda \in { 1\over \xi} [-1/3,0)$ makes $Det$ zero, things
become a little more complex since in this case, we can choose the parameters
$k_1,k_2,r_1(k),r_2(k)$ properly to have nonzero $B_{e,o}^{(1,2)}$. This
opens a door to new solutions. Let us discuss it in more detail. 
First, we choose the parameters such that $B_{e,o}^{(1,2)}=0$. In this case,
we have two solutions given by the same form as \eref{discrete_1}. 

We need two more solutions and they can be found by choosing nonzero 
values of $B_{e,o}^{(1,2)}\neq 0$.
Let us discuss this case in more detail. To find these
two more solutions we need to generalize our 
method a little bit. The idea is that
when we had $\int dk f(k) \ket{k}=0$ before, in general we should require
\[
f(k) = \sum_{k_i} \delta(k-k_i) r_i(k),~~~~~r_i(k_i)=0.
\]
Since the equation is linear, if we find a single solution, 
we can get the general one by
linear combination. This method
works well for the analysis in the previous sections. 
However, in the case at hand, such a
simplification does not work. Let us explain a little bit.
Assume there is a solution for  $k_1=k_2=k_0$, $r_1(k)=D_1 (k-k_0)$
and $r_2(k)=D_2 (k-k_0)$ with $\lambda=M(k_0)/\xi$. Consistency requires
that
\[
{d_{oe}A_{ee} \over 1-d_{oo} A_{oo}} = -{ B_{e,k_0} D_1 \over B_{o,k_0} D_2}
= { B_{e,k_0} D_2 \over B_{o,k_0} D_1},~~~~\Rightarrow~~~D_1^2=-D_2^2
\]
which have solutions only when $D_1=\pm i D_2$ and are perfectly
consistent since $d_{oe}$ is a purely imaginary number.
Now the crucial thing is to check if
\[
{ d_{oe}A_{ee} \over 1-d_{oo} A_{oo}} = -i{ B_{e,k_0} \over B_{0,k_0}}~~~~
\mbox{or}~~~{ d_{oe}A_{ee} \over 1-d_{oo} A_{oo}} = i{ B_{e,k_0} \over
B_{0,k_0}} 
\]
for the given $\lambda$ which makes $Det=0$. We do not have an
analytic proof to show this is not true, but it is easy to check
numerically and was found that it indeed does not hold. In other
words, if we keep only a single term we do not have any solution. 

In our case, the generalization is obvious: we just replace
$\delta(k-k_1) r_1(k)$ by $\delta(k-k_1) r_1(k)+\delta(k+k_1)
\tilde{r}_1(k)$. This then modifies \eref{mateigen} to
\be
\label{modified}
\left( \begin{array}{cccc}  1-d_{ee} A_{ee}~~ & 0 & 0 & -d_{oe} A_{ee} \\
  0 & 1-d_{oo} A_{oo}~~ & -d_{oe} A_{oo} & 0 \\
  0 &  d_{oe} A_{ee}  & 1- d_{ee} A_{ee}~~ & 0 \\
  d_{oe} A_{oo} & 0 & 0 & 1-d_{oo} A_{oo}
\end{array} \right)
\left( \begin{array}{l} {\cal C}^{(1)}_e \\ {\cal C}^{(1)}_o \\
  {\cal C}^{(2)}_e \\{\cal C}^{(2)}_o  \end{array} \right)
=\left( \begin{array}{l} B_{e,k_0}D_1+B_{e,-k_0}\widetilde{D}_1 \\ 
B_{o,k_0}D_1+B_{o,-k_0}\widetilde{D}_1 \\
 B_{e,k_0}D_2+B_{e,-k_0}\widetilde{D}_2 \\
B_{o,k_0}D_2+B_{o,-k_0}\widetilde{D}_2
 \end{array} \right).
\ee

Now consistency requires
\[
{ d_{oe}A_{ee} \over 1-d_{oo} A_{oo}}
=-{B_{e,k_0}D_1+B_{e,-k_0}\widetilde{D}_1
\over B_{o,k_0}D_2+B_{o,-k_0}\widetilde{D}_2}=
{B_{e,k_0}D_2+B_{e,-k_0}\widetilde{D}_2 \over
B_{o,k_0}D_1+B_{o,-k_0}\widetilde{D}_1 }.
\]
Using $B_{e,k_0}=B_{e,-k_0}$ and $B_{o,k_0}=B_{o,-k_0}$ we can
solve 
\[
D_2={1\over 2}[(\rho-{1\over \rho})D_1-(\rho+{1\over
\rho})\widetilde{D}_1],
~~~~~
\widetilde{D}_2={1\over 2}[(\rho-{1\over \rho})D_1-(\rho-{1\over
\rho})\widetilde{D}_1],
\]
where we have defined $\rho={ d_{oe}A_{ee} \over 1-d_{oo} A_{oo}}
{B_{o,k_0} \over B_{e,k_0}}$. Notice that we have two independent
parameters $D_1,\widetilde{D}_1$ which give us the missing two
eigenstates. However, the most convenient way is to choose 
$$
D_{i;+}=D_i+\widetilde{D}_i,~~~~~D_{i;-}=D_i-\widetilde{D}_i.
$$
Then \eref{modified} becomes
\[
\left( \begin{array}{cccc}  1-d_{ee} A_{ee}~~ & 0 & 0 & -d_{oe} A_{ee} \\
  0 & 1-d_{oo} A_{oo}~~ & -d_{oe} A_{oo} & 0 \\
  0 &  d_{oe} A_{ee}  & 1- d_{ee} A_{ee}~~ & 0 \\
  d_{oe} A_{oo} & 0 & 0 & 1-d_{oo} A_{oo}
\end{array} \right)
\left( \begin{array}{l} {\cal C}^{(1)}_e \\ {\cal C}^{(1)}_o \\
  {\cal C}^{(2)}_e \\{\cal C}^{(2)}_o  \end{array} \right)
=\left( \begin{array}{l}  D_{1,+}B_{e,k_0} \\ D_{1,-} B_{o,k_0}\\
 \rho  D_{1,-}B_{e,k_0}  \\ -{1\over \rho} D_{1,+}B_{o,k_0}
 \end{array} \right),
\]
from which we have two solutions
\be
\left( \begin{array}{l} {\cal C}^{(1)}_e \\ {\cal C}^{(1)}_o \\
  {\cal C}^{(2)}_e \\{\cal C}^{(2)}_o  \end{array} \right)
=\left( \begin{array}{l} 0 \\ 0 \\ 0 \\ -{D_{1,+}B_{e,k_0} 
\over  d_{oe}A_{ee}} \end{array} \right),~~~~~~
\left( \begin{array}{l} {\cal C}^{(1)}_e \\ {\cal C}^{(1)}_o \\
  {\cal C}^{(2)}_e \\{\cal C}^{(2)}_o  \end{array} \right)
=\left( \begin{array}{l} 0 \\ {\rho  D_{1,-}B_{e,k_0} \over
  d_{oe}A_{ee}} \\0 \\ 0 \end{array} \right).
\ee

Our final eigenvectors then become
\be
v_{+} =\left( \begin{array}{l} 
-{\sqrt{2b} \xi \over 2 (\lambda \xi-1)}
{B_{e,k_0}D_{1,+}\over  A_{ee}}\\  \\
-{ B_{e,k_0}D_{1,+} \over  A_{ee}(\lambda-{M \over \xi})} \ket{v_e}
-{\xi D_{1,+} \over 2\frac{dM}{dk}|_{k_0}} (\ket{k_0}-\ket{-k_0})
\\  \\ 0 \\  \\
-{d_{oo} B_{e,k_0}D_{1,+} \over d_{oe} A_{ee}(\lambda-{M \over \xi})} \ket{v_o}
+{\xi {1\over \rho}D_{1,+} \over 2\frac{dM}{dk}|_{k_0}}(\ket{k_0}+\ket{-k_0})
\end{array} \right)
\ee
and
\be
v_{-}=\left( \begin{array}{l} 0 \\ \\
{\rho d_{oo} B_{e,k_0}D_{1,-} \over d_{oe} A_{ee}(\lambda-{M \over \xi})} \ket{v_o}
-{\xi D_{1,-} \over 2\frac{dM}{dk}|_{k_0}}(\ket{k_0}+\ket{-k_0}) \\ \\
{-\rho\sqrt{2b} \xi \over 2 (\lambda \xi-1)}
{B_{e,k_0}D_{1,-}\over  A_{ee}}\\  \\
{-\rho B_{e,k_0}D_{1,-} \over  A_{ee}(\lambda-{M \over \xi})} \ket{v_e}
-{\xi \rho D_{1,-} \over 2\frac{dM}{dk}|_{k_0}}(\ket{k_0}-\ket{-k_0})
\end{array} \right) \,.
\ee

\section{Level truncation analysis}

Here we want to compare some of our predictions with the results of level
truncation. First, we do observe a continuous spectrum in the range
$\xi^{-1} [-{1\over 3}, 0)$. Next, we want to check our analytical
expressions for
the eigenvalues in the interval $(0, \xi^{-1})$. For this, we show in the
following table these rescaled eigenvalues $\xi \lambda_1$ and $\xi
\lambda_2$
(defined such that $\lambda_1 \leq \lambda_2$), for $b=1$ and for various
values
of $B$, to levels 5, 20 and 100. We compare these results to the exact
values
calculated from \eref{transc}, shown in the last column.

\begin{center}
\vspace{6pt}
\begin{tabular}{|c||c|c|c||c|}  \hline
 & Level 5  & Level 20  & Level 100  & Exact value \\  \hline \hline
$\xi \lambda_1(B = 0.001)$  & 0.394832 & 0.397808 & 0.397969 & 0.397976
\\ \hline
$\xi \lambda_2(B = 0.001)$  & 0.408028 & 0.410125 & 0.410283 & 0.41029
\\ \hline \hline
$\xi \lambda_1(B = 0.01)$  & 0.340706 & 0.343376 & 0.343537 & 0.343544
\\ \hline
$\xi \lambda_2(B = 0.01)$  & 0.463391 & 0.465736 & 0.465892 & 0.465899
\\ \hline \hline
$\xi \lambda_1(B = 0.1)$  & 0.0353826 & 0.036406 & 0.0364938 & 0.0364979
\\ \hline
$\xi \lambda_2(B = 0.1)$  & 0.83877 & 0.839611 & 0.839684 & 0.839688
\\ \hline \hline
$\xi \lambda_1(B = 0.5)$  &$-0.000444917$ & $-3.03731\cdot 10^{-15}$ &
$7.39318\cdot 10^{-18}$ & $4.08988\cdot 10^{-69}$
\\ \hline
$\xi \lambda_2(B = 0.5)$  & 0.989926 & 0.98998 & 0.989986 & 0.989986
\\ \hline
\end{tabular}
\end{center}

We see a remarkable agreement between the level truncation and
our analytical values. As was already observed in \cite{mprime}, level
truncation converges very fast for the isolated eigenvalues.

This also illustrates our discussion in section \ref{zeros01} on how turning on a
$B$-field
breaks the degeneracy between $\lambda_1$ and $\lambda_2$. Indeed we see
that
$\xi \lambda_2$ goes to one as $B$ is increased, whereas $\xi \lambda_1$
goes to zero.
This last convergence being very fast, as can be measured from the
asymptotic
expansions of our analytical expressions \cite{mprime}.

Next we want to check our formulas \eref{eigenc1} and 
\eref{eigenc2}\footnote{The vectors 
$\left( \lambda - M/\xi \right)^{-1} |v_{e,o}\rangle$ can be calculated 
from our formulas for the generating functions written down in \cite{mprime}} 
for the eigenvectors in the continuous 
spectrum $v^e_{1, \lambda}$, $v^o_{1, \lambda}$, $v^o_{2, \lambda}$ and 
$v^e_{2, \lambda}$. Here some care is needed: Due to the 
block-matrix form \eref{M_11} of $({\cal M})^{11}$, its eigenvalues in the 
level truncation 
will always be twice exactly degenerate; indeed if (in block notation) 
$\left(\begin{array}{c} a \\ b \end{array} \right)$ is an eigenvector, then so is 
$\left(\begin{array}{c} b \\ -a \end{array} \right)$. This degeneracy is 
undesirable because the 
numerical algorithm which finds the level-truncated eigenvectors will, for 
each eigenvalue, give two eigenvectors which will be in a mixed state. In other 
words, for some eigenvalue $\lambda$, the algorithm will output two vectors 
$w_1 = a_1 v^e_{1, \lambda} + a_2 v^o_{1, \lambda} + a_3 v^o_{2, \lambda} + 
a_4 v^e_{2, \lambda}$ and $w_2 = b_1 v^e_{1, \lambda} + b_2 v^o_{1, \lambda} 
+ b_3 v^o_{2, \lambda} + b_4 v^e_{2, \lambda}$, and we will have no control 
on the parameters $a_i$ and $b_i$. We can remedy\footnote{The reader 
might wonder 
why we don't simply check our exact eigenvectors by multiplying them 
by the level truncated $({\cal M})^{11}$. The problem is that this procedure does 
not give very good results in the level truncation.}  to this by artificially 
breaking 
this degeneracy in the following way: We consider 
$({\cal M})^{11}_\epsilon \equiv ({\cal M})^{11} + \epsilon J$, where 
$J \equiv \left( \begin{array}{cc} 1 & 0 \\ 0 & -1 \end{array} \right)$, and 
$\epsilon$ is a small number (we will typically choose $\epsilon = 
10^{-7}$. We observe that adding this small perturbation to $({\cal M})^{11}$ 
breaks 
the degeneracy in the level truncation. moreover, $J$ commutes with 
$\left( \begin{array}{cc} C & 0 \\ 0 & -C \end{array} \right)$, we thus expect 
the level-truncated eigenvectors to be of the form 
$\left( \begin{array}{c} {\rm even} \\ {\rm odd} \end{array} \right)$ which we 
will call {\em even}, or 
$\left( \begin{array}{c} {\rm odd} \\ {\rm even} \end{array} \right)$ which we 
will call {\em odd}. If we 
further constraint all eigenvectors to be of the form 
$\left( \begin{array}{c} {\rm real} \\ {\rm imaginary} \end{array} \right)$, we 
expect the even level-truncated eigenvectors $w^e_\lambda$ to be close to 
some linear combination 
$a_1 v^e_{1, \lambda} + a_2 v^e_{2, \lambda}$ with $a_1$ and $a_2$ being real 
parameters. Similarly, we expect the odd level-truncated eigenvectors 
$w^o_\lambda$ to be close 
to some linear combination 
$a_1 v^o_{1, \lambda} + a_2 v^o_{2, \lambda}$.

In the next tables, we consider 4 eigenvectors of the level-truncated 
matrix $({\cal M})^{11}_\epsilon$ with $b=1$, $B=0.2$, $\epsilon = 10^{-7}$ 
and we work at level $L=100$. We find the parameters $a_1$ and $a_2$ by 
minimizing 
$$
{\rm err}(a_1, a_2) = {\left( \sum_{i=0}^5{|(a_1 v^{e,o}_{1, \lambda} + 
a_2 v^{e,o}_{2, \lambda} - w^{e,o}_\lambda)_i|^2} + 
\sum_{i=L+1}^{L+6}{|(a_1 v^{e,o}_{1, \lambda} + 
a_2 v^{e,o}_{2, \lambda} - w^{e,o}_\lambda)_i|^2} \right)^{1 \over 2} \over 
\|a_1 v^{e,o}_{1, \lambda} + a_2 v^{e,o}_{2, \lambda}\|} \,,
$$
where we have chosen, for simplicity, to compare only the 6 first components of 
the upper block and the 6 first components of the lower block (remember that each 
block has size $L+1$). Also, all the vectors appearing in the above
formula are normalized to unit norm before being plugged in the
formula. Here are the results of this fit for two different 
eigenvalues which appear in the level truncation.

\begin{center}
\vspace{6pt}
\begin{tabular}{|c||c|c|}  \hline
$\lambda = -0.110388$, even vector& Level truncation  & Exact value \\  \hline \hline
component 0 & 0.158253 & 0.158271
\\ \hline
component 2 & $-0.314559$ & $-0.314559$
\\ \hline
component 4 & 0.206995 & 0.206947
\\ \hline
component $L+2$ & $-0.583525 \,i$ & $-0.583576 \,i$
\\ \hline
component $L+4$ & $0.313203 \,i$ & $0.313186 \,i$
\\ \hline
component $L+6$ & $-0.228656 \,i$ & $-0.22858 \,i$
\\ \hline \hline
$a_1$ & \multicolumn{2}{|c|}{$-0.0289756$} 
\\ \hline
$a_2$ & \multicolumn{2}{|c|}{0.884282} 
\\ \hline
err & \multicolumn{2}{|c|}{$0.0107\%$} 
\\ \hline
\end{tabular}
\end{center}

\begin{center}
\vspace{6pt}
\begin{tabular}{|c||c|c|}  \hline
$\lambda = -0.110388$, odd vector& Level truncation  & Exact value \\  \hline \hline
component 1 &0.583524  & 0.583575
\\ \hline
component 3 & $-0.313202$ & $-0.313185$
\\ \hline
component 5 & 0.228656 & 0.228578
\\ \hline
component $L+1$ & $-0.158253 \,i$ & $-0.158271 \,i$
\\ \hline
component $L+3$ & $ 0.314561 \,i$ & $ 0.314561\,i$
\\ \hline
component $L+5$ & $-0.206996 \,i$ & $-0.206948 \,i$
\\ \hline \hline
$a_1$ & \multicolumn{2}{|c|}{$-0.0289793$} 
\\ \hline
$a_2$ & \multicolumn{2}{|c|}{0.884282} 
\\ \hline
err & \multicolumn{2}{|c|}{$0.0107\%$} 
\\ \hline
\end{tabular}
\end{center}

\begin{center}
\vspace{6pt}
\begin{tabular}{|c||c|c|}  \hline
$\lambda = -0.0716535$, even vector& Level truncation  & Exact value \\  \hline \hline
component 0 & 0.0342766 & 0.0342883
\\ \hline
component 2 & 0.280389 & 0.280447
\\ \hline
component 4 & $-0.260653$ & $-0.260619$
\\ \hline
component $L+2$ & $-0.107005 \,i$ & $-0.107033 \,i$
\\ \hline
component $L+4$ & $0.201759 \,i$ & $0.201769 \,i$
\\ \hline
component $L+6$ & $ -0.185215\,i$ & $-0.185146 \,i$
\\ \hline \hline
$a_1$ & \multicolumn{2}{|c|}{0.673384} 
\\ \hline
$a_2$ & \multicolumn{2}{|c|}{$-0.100719$} 
\\ \hline
err & \multicolumn{2}{|c|}{$0.0102\%$} 
\\ \hline
\end{tabular}
\end{center}

\begin{center}
\vspace{6pt}
\begin{tabular}{|c||c|c|}  \hline
$\lambda = -0.0716536$, odd vector& Level truncation  & Exact value \\  \hline \hline
component 1 & 0.107009 & 0.107037
\\ \hline
component 3 & $-0.20176$ & $-0.20177$
\\ \hline
component 5 & 0.185216& 0.185147
\\ \hline
component $L+1$ & $-0.0342776 \,i$ & $ -0.0342892\,i$
\\ \hline
component $L+3$ & $-0.280388 \,i$ & $-0.280446 \,i$
\\ \hline
component $L+5$ & $0.260652 \,i$ & $0.260619 \,i$
\\ \hline \hline
$a_1$ & \multicolumn{2}{|c|}{$0.673384$} 
\\ \hline
$a_2$ & \multicolumn{2}{|c|}{$-0.100714$} 
\\ \hline
err & \multicolumn{2}{|c|}{$0.0102\%$} 
\\ \hline
\end{tabular}
\end{center}

We see that the errors are of the order of $0.01\%$. This is therefore a very 
good test of the validity of our formulas. 

At last, let us do the same kind of test for one discrete eigenvalue in order to check
\eref{discrete_1}. 
With the same parameters as before, we have the two 
following discrete eigenvalues: $\lambda_1 = 0.0000900753$ and $\lambda_2 = 0.366192$. 
In the following two tables, we compare the eigenvectors related to $\lambda_2$. 
Here, half of the degeneracy is already broken by the $B$-field. Adding the perturbation 
will further break the degeneracy completely, and we don't need to use a fit.

\begin{center}
\vspace{6pt}
\begin{tabular}{|c||c|c|}  \hline
$\lambda = 0.366192$, even vector& Level truncation  & Exact value \\  \hline \hline
component 0 & 0.97589 & 0.975946
\\ \hline
component 2 & 0.0115766 & 0.0115736
\\ \hline
component 4 & $-0.00544373$ & $-0.00543968$
\\ \hline
component $L+2$ & $ 0.21255\,i$ & $ 0.212556\,i$
\\ \hline
component $L+4$ & $ -0.0404268\,i$ & $ -0.0404213\,i$
\\ \hline
component $L+6$ & $ 0.0186439\,i$ & $ 0.0186363\,i$
\\ \hline \hline
err & \multicolumn{2}{|c|}{$0.0026\%$} 
\\ \hline
\end{tabular}
\end{center}

\begin{center}
\vspace{6pt}
\begin{tabular}{|c||c|c|}  \hline
$\lambda = 0.366192$, odd vector& Level truncation  & Exact value \\  \hline \hline
component 1 & 0.212551 & 0.212556
\\ \hline
component 3 & $-0.0404268$ & $-0.0404213$
\\ \hline
component 5 & 0.0186439 & 0.0186363
\\ \hline
component $L+1$ & $0.97589 \,i$ & $ 0.975946\,i$
\\ \hline
component $L+3$ & $ 0.0115766\,i$ & $ 0.0115737\,i$
\\ \hline
component $L+5$ & $ -0.00544373\,i$ & $ -0.00543973\,i$
\\ \hline \hline
err & \multicolumn{2}{|c|}{$0.0141\%$} 
\\ \hline
\end{tabular}
\end{center}

Again, there is a remarkable agreement between numerical and analytical results. 
These calculation also show that level truncation gives very precise results for 
the eigenvectors; this observation was already made in \cite{Spec}.
\section{Conclusions and Discussions}

We analytically solved the problem of finding all eigenvalues and
eigenvectors of the Neumann matrix ${\cal M}^{11}$ of the bosonic
star-product in the matter  
sector including zero-modes and with a nontrivial $B$-field background.
In particular we see, as was already observed in \cite{Bonora}, that
turning  on a $B$-field does not seem to obstruct the possibility of
finding exact solutions. 

Starting from the spectrum of ${\cal M}^{11}$ with $B=0$, we found
that turning on the $B$-field has three effects on the spectrum:
\begin{enumerate}
\item It shrinks the whole spectrum by $\xi = 1 + (2 \pi \alpha' B)^2$.
\item It splits the discrete eigenvalue into two different ones
(Zeeman effect), still in the 
	range $(0, {1\over \xi})$.
\item As we are considering two spatial directions, the eigenvalues in 
$(-{1\over 3\xi}, 0)$ are now four times degenerate. Surprisingly,
when $B \neq 0$, the eigenvalue $-{1\over 3\xi}$ is only doubly
degenerate. Also the isolated eigenvalues in the range $(0, {1\over
\xi})$ are both doubly degenerate. 
\end{enumerate}
In the discussion of
$Det$ we also found some discrete values 
inside the interval $[-{1\over 3 \xi}, 0)$, depending on the 
parameters $b$ and $B$, for which 
the expressions of the eigenvectors differ
from those of the continuous eigenvectors. However,
it is merely an artifact 
of the basis in which we are expanding our eigenvectors, and if we
renormalize the expression by a proper factor it will be continuous
in the whole region $[{-1\over 3 \xi},0)$\footnote{We 
thank D.~Belov for discussing this issue.}. Furthermore,
using the same trick as in \cite{Spec} and \cite{mprime}, it is an 
obvious task to calculate the spectra of ${\cal M}^{12}$ and ${\cal M}^{21}$ 
from the spectrum of ${\cal M}^{11}$ since the three of them satisfy
the same algebra relations \cite{Bonora}.

The physical meaning of the discrete eigenvalues is still
mysterious. It seems  
however that, unlike the continuous ones, the discrete eigenvectors
are $\ell^2({\bf N})$-normalizable; this hints us that they are 
particularly interesting\footnote{We thank W.~Taylor and 
B.~Zwiebach for a discussion about this point.}. It thus seems very 
important to 
try to understand the meaning of these eigenvectors. 
We can actually get some clue about it by considering the large
$B$-field limit
\cite{Witten_B, Schnabl_B}. Indeed if we write the expansion of 
${\cal M}^{11}$ in powers of ${1\over B}$, using \eref{omegaxi} 
and \eref{M_11}, we find, up to order two:
\bea
&& \xi {\cal M}^{11} = \left(\begin{array}{ll|ll} 
1 & 0 & 0 & 0
\\
0 & M & 0 & 0
\\ \hline
0 & 0 & 1 & 0
\\
0 & 0 & 0 & M
\end{array} \right) + 
\nonumber \\
&& + {1 \over B} {i \over \sqrt{3} \pi^2} 
\left(\begin{array}{ll|ll} 
0 & 0 & 0 & -\sqrt{2b} \langle v_o| 
\\ 
0 & 0 & -\sqrt{2b} |v_o \rangle & 
-2 (|v_e\rangle \langle v_o| + |v_o\rangle \langle v_e|) 
\\ \hline 
0 & \sqrt{2b} \langle v_o| & 0 & 0 
\\ 
\sqrt{2b} |v_o \rangle & 
2 (|v_e\rangle \langle v_o| + |v_o\rangle \langle v_e|) & 0 & 0
\end{array} \right) + 
\nonumber \\ 
&& + {1 \over B^2} {- \beta \over 2 \pi^4} 
\left(\begin{array}{ll|ll} 
b & \sqrt{2b} \langle v_e| & 0 & 0
\\
\sqrt{2b} |v_e \rangle & 2 (|v_e\rangle \langle v_e| - 
|v_o\rangle \langle v_o|) & 0 & 0
\\ \hline
0 & 0 & b & \sqrt{2b} \langle v_e|
\\
0 & 0 & \sqrt{2b} |v_e \rangle & 2 (|v_e\rangle \langle v_e| - 
|v_o\rangle \langle v_o|)
\end{array} \right) + {\cal O}\left( \left( {1\over B} \right)^3
\right) \,. \label{limitcase}
\eea
If we look only to order zero, we see immediately that the spectrum of 
$\xi {\cal M}^{11}$ is two copies of the spectrum of $M$ and 
another two eigenvectors with
the eigenvalue one. The one with unit eigenvalue obviously is one of our
discrete eigenvalues (the other one is sent to zero as $B \rightarrow
\infty$). On the other hand, we know from \cite{Witten_B, Schnabl_B},
that in the large $B$-field limit, the star-algebra factorizes into a
zero-momentum star-algebra (represented here by the spectrum of $M$)
and one non-commutative algebra (which must therefore be related to the
discrete eigenvalue) corresponding to the momentum of the
string. Putting it another way, the discrete eigenvalues should
be related to the momentum of the string. 
All these discussions of the limiting case 
give us a hint on the physical meaning of these discrete 
eigenvectors we met in \cite{mprime} and here. 

There is another thing we can observe from equation
(\ref{limitcase}). In the limit of large $B$, 
the mixing of the two spatial directions
happens only at the first order. In other words, comparing with
the non-commutativity from the star-product in string field theory
(both the zero mode and non-zero modes), the non-commutativity from
the $B$-field in the spatial directions is small at large $B$ limit.
This may be a little of surprising and deserves better understanding. 

There are some immediate follow-up works, for example, 
to calculate the spectrum of the Neumann matrices in the 
ghost sector in the light of 
\cite{Moyal}. This will clarify some subtle points, such as the
infinite coefficients mentioned therein.
Another interesting direction is to discuss the
Moyal product corresponding to the case of $B$ fields. From our
eigenvectors we see a mixing of even (odd) state in first direction 
with the odd (even) in the second. This pattern tells us how
the noncommutativities from the star-product and from the $B$-field mix
up. Understanding this mix-up will help us better comprehend the
noncommutativity.

\section*{Acknowledgements}

We extend our sincere gratitude to B.~Zwiebach for useful discussions
and proofreading of the paper. We would also like to thank I.~Ellwood, 
M.~Mari\~{n}o and R.~Schiappa
for many stimulating conversations. 
Many thanks also to D.~Belov for interesting correspondence and 
for showing us his unpublished work on the spectrum of $M'$, 
and to W.~Taylor for discussions.
This Research was supported in part by
the CTP and LNS of MIT and the U.S. Department of Energy 
under cooperative research agreement \# DE-FC02-94ER40818.

%%%%%%%%%%%%%%%%%%%%%%%%%%%%%%%%%%%%%%%%%%%%%%%%%%%%%%%%%%%%%
%%%%%%%%%%%%%%%%%%%%%%%%%%%%%%%%%%%%%%%%%%%%%%%%%%%%%%%%%%%%%

\bibliographystyle{JHEP}

\end{document}